\documentclass[11pt]{article} 

\usepackage[utf8]{inputenc} 

\usepackage{geometry} 
\usepackage{graphicx,url} 

\usepackage{booktabs} 
\usepackage{array} 
\usepackage{paralist} 
\usepackage{verbatim} 
\usepackage{subfig} 
\usepackage{amssymb}
\usepackage{amsmath}
\usepackage{textcomp} 

\usepackage{fancyhdr} 
\usepackage{sectsty}
\allsectionsfont{\sffamily\mdseries\upshape} 

\usepackage[nottoc,notlof,notlot]{tocbibind} 
\usepackage[titles,subfigure]{tocloft} 

\newcommand{\com}[1]{\textnormal{#1}}

\title{\textbf{Self-gravitating dark matter gets in shape}\vspace{2ex}}
\author{
\textbf{Jenny Wagner} \\[1ex] 
j.wagner@uni-heidelberg.de \\
Universit\"at Heidelberg, ZAH, ARI, \\ 
M\"onchhofstr.~12--14, 69120 Heidelberg, Germany \\ 
\url{https://www.zah.uni-heidelberg.de/staff/jwagner/}}

\begin{document}
\maketitle
\begin{center}
\textit{Essay received an Honorable Mention \\ in the 2020 Essay Competition of the Gravity Research Foundation.}
\end{center}
\vspace{1ex}
\begin{figure}[h!]
\begin{center}
\includegraphics[width=0.6\textwidth]{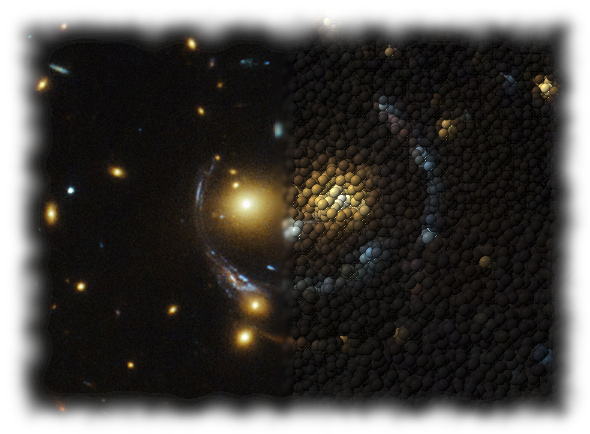}
\end{center}
\end{figure}
\abstract{
\noindent 
In our current best cosmological model, the vast majority of matter in the Universe is dark, consisting of yet undetected, non-baryonic particles that do not interact electro-magnetically. 
So far, the only significant evidence for dark matter has been found in its gravitational interaction, as observed in galaxy rotation curves or gravitational lensing effects.
The inferred dark matter agglomerations follow almost universal mass density profiles that can be reproduced well in simulations, but have eluded an explanation from a theoretical viewpoint.
Forgoing standard (astro-)physical methods, I show that it is possible to derive these profiles from an intriguingly simple mathematical approach that directly determines the most likely spatial configuration of a self-gravitating ensemble of collisionless dark matter particles. 
}

\newpage
\noindent
\textbf{Ubiquitous universality}
\\[2ex]
Dark matter may be a mysterious form of matter, yet, its gravitational interaction can be reconstructed well in numerous examples like tracing the rotation curves of stars in a galaxy \cite{bib:Blok},\cite{bib:Rubin2}, or observing the gravitational lensing effect of galaxies and galaxy clusters whose masses distort light bundles on their way from light-emitting background objects to us \cite{bib:Meylan}.
These observations show a high uniformity in the inferred dark matter mass density profiles.
A small set of heuristic, parametric models fit a wide range of galaxy and galaxy cluster mass densities.
$N$-body simulations that emulate the process of dark matter agglomeration with gravity as the only interaction corroborate these findings \cite{bib:NFW1}, \cite{bib:NFW2}.
But they also reveal deviations from universal mass density profiles with increasing resolution in length and mass scales \cite{bib:NFW3}.

Numerous ideas have been developed to derive the shape of dark matter mass density profiles from statistical mechanics as equilibrium configurations with maximum entropy in phase-space or energy-space \cite{bib:Hjorth}, \cite{bib:Salvador}.
Although they greatly enhance our understanding, some fundamental questions remain. 
For instance, how is the innermost part of a dark matter structure, we call ``dark matter halo", shaped? 
Why does the outer halo region on galaxy scale show a steeper decrease in mass density than its galaxy-cluster-scale counterpart? 
Why does universality dissolve with increasing resolution? 
Why do halo shapes seem to be independent of their mass accretion history and the background cosmology?

As I show in the following and further detail in \cite{bib:Wagner_halo}, all these questions find an answer in a simple mathematical approach that reverse engineers dark matter mass density profiles. 
Contrary to standard methods, it separates the morphological description of a halo from its dynamics and focuses on the spatial distribution of dark matter particles.
The particle interactions are phenomenologically modelled by the mean gravitational field they generate themselves.
This minimalistic approach \com{to describe the mass density of dark matter halos} does not require any definition of phase-space configurations, entropy, or the usage of the particle velocities.
\\[4ex]
\noindent
\textbf{Convincing characterisations}
\\[2ex]
\com{To find a macroscopic, effective model that describes the continuous mass density of a dark matter halo at one instant in cosmic time,} we \com{set up a statistical ensemble of} a finite amount of $n_\mathrm{p}$ dark matter particles \com{that forms this dark matter halo}.
\com{Any occurring divergences in the modelling are prevented by assuming that these particles always keep a finite minimum distance and that the halo they shape is of finite volume.}
Furthermore, we restrict our model to identical, \com{identically distributed} particles and spherical halo volumes, such that all equations are analytically solvable and the concept is clearly recognisable.

\com{Taking the definition of a mass density literally, the morphological description of the halo at one moment in time can be completely determined by observing the momentary positions of all massive particles. 
When introducing conservation laws, like energy conservation, particle positions become related to momenta and observations of any combination of positions or momenta can be employed to describe the spatial halo morphology.
Yet, as we will show below, it is not necessary to introduce these additional relations and constraints to understand the shape of dark matter halos.
In fact, the instantaneous dark matter halo description that is only based on particle positions resolves long-standing issues that have prevented derivations of halo mass densities from standard statistical mechanics approaches. 
Among them are the problem of a missing equation of state of dark matter, or more fundamental questions whether self-gravitational ensembles can be described using Boltzmann-Gibbs statistics, see e.g. \cite{bib:Hjorth} and \cite{bib:Levin} for further details. 
\\
As many works have shown, relaxation processes based on collisions of dark matter particles are too slow to reproduce the large number of relaxed, quasi-stable dark matter halos inferred from observational data, see e.g. the recent work of \cite{bib:Fouvry} for estimates about the time scales on which collisionless, two-particle and three-particle collisional relaxation occur. 
On short time scales compared to the age of the universe, collisionless relaxation dominates over collisional relaxation for large ensembles of particles as derived in \cite{bib:Braun}. 
Due to this dominance of collisionless relaxation for time scales and particle numbers relevant to galaxy-scale and galaxy-cluster scale dark matter halos, we also assume that the particles in the ensemble are collisionless, i.e.~independent of each other and only interacting with the mean gravitational field that they generate.
Consequently, each particle of the ensemble follows the same probability density function (PDF) $p(r)$ to be located at a specific radial position $r$ inside the halo volume with maximum radius $r_\mathrm{max}$.
}
Taking into account that the mean gravitational field of the ensemble is generated by applying Newton's scale-free gravitation to all particle pairs, we \com{assume that the resulting particle distribution that builds the mean gravitational field is also scale-free. 
We therefore parametrise $p(r)$ as a general} power-law \com{PDF} 
\begin{equation}
p(r) = N(\alpha,r_\sigma,r_\mathrm{max}) \left( \dfrac{r}{r_\sigma} \right)^{-\alpha} \;, \quad  \alpha \ge 0 \;,
\label{eq1}
\end{equation}
with power-law index $\alpha$, and scale-radius $r_\sigma$, introduced to obtain dimensionless quantities.
$N(\alpha,r_\sigma,r_\mathrm{max})$ normalises $p(r)$, such that the probability of finding the particle in the halo volume equals one.
\com{This normalisation implies that a halo around $r=0$ exists and the particles of the ensemble are assigned to it. Thus, the divergence at $r=0$ is avoided, because $p(r)$ is interpreted as the PDF for a particle to be at radius $r$ given that it belongs to the halo existing around a most-bound particle at $r=0$, which usually defines the centre of a halo in simulations.}
\com{Based on these prerequisites,} the joint probability density function to find the ensemble in a specific spatial configuration is given by multiplying the $p(r_i)$ for all independent particles $i=1,...,n_\mathrm{p}$. 

\com{Having set up a description for the particle ensemble in terms of the joint PDF, we investigate which power-law indices belong to distributions of particles that are very likely to occur. 
When analysing a set of dark matter halos, we expect the most likely shape to have the highest occurence.}
\com{The extremum} configurations with respect to $\alpha$ \com{are obtained by taking the derivative of} the logarithm of the joint \com{PDF with respect to $\alpha$ and setting it to zero}
\begin{equation}
\dfrac{\partial_\alpha N(\alpha,r_\sigma,r_\mathrm{max})}{N(\alpha,r_\sigma,r_\mathrm{max})} - \dfrac{1}{n_\mathrm{p}} \sum \limits_{j=1}^{n_\mathrm{p}} \ln \left( \dfrac{r_j}{r_\sigma} \right) \stackrel{!}{=} 0 \;.
\label{eq2}
\end{equation}
\com{While deriving the joint PDF with respect to $\alpha$, $r_\sigma$ and $r_\mathrm{max}$ are kept as fixed parameters.}
We note that $\alpha$ enters \com{Equation~\ref{eq2}} via the normalisation, i.~e.~through the assumed halo geometry and its volume defined in $N$. 
\com{The choice of $r_\mathrm{max}$ sets the length scale of interest to represent the particle ensemble as a dark matter halo.
As we will see below in Equation~\ref{eq3}, the resulting $\alpha$ is independent of $r_\sigma$, corroborating its role as a mere auxiliary scaling radius in Equation~\ref{eq1}.}

The sum-term containing the particle number and distribution of the ensemble accounts for resolution effects.
Equation~\ref{eq2} \com{thus links the ``microscopic" ensemble of individual particles, represented by the sum term to its ``macroscopic" effective representation in terms of a dark matter halo parametrised by $\alpha$ and $r_\mathrm{max}$, represented in the normalisation $N$.
The equation} is invariant for distinguishable and indistinguishable particles because the respective prefactor in the joint \com{PDF} is independent of $\alpha$. 

Before solving Equation~\ref{eq2} to obtain $\alpha$ for different physical approximations, we need to derive the continuous halo mass density, $\rho(r)$, from the single-particle probability density function (Equation~\ref{eq1}). 
This is easily achieved, because, the number density $n(r)$ for our spherical halo of collisionless particles is defined as the phase-space probability density function for a single particle after marginalising out the velocity. 
If we interpret Equation~\ref{eq1} as this spatial part of the single-particle phase-space probability density and multiply $n(r)$ by the mass of a particle $m$ we arrive at
\begin{equation}
n(r) = n_\mathrm{p} \, p(r) \quad \Rightarrow \quad \rho(r) = m n(r) = m n_\mathrm{p} \, p(r)\;.
\end{equation}
Hence, $\rho(r)$ obeys the same power-law of Equation~\ref{eq1}, which means that the slope of the mass density profiles can be directly related to $\alpha$ for the extremum configurations of the particle ensemble determined by Equation~\ref{eq2}.  

\com{Inserting Equation~\ref{eq1} into Equation~\ref{eq2}, we obtain}
\begin{equation}
\dfrac{1}{\alpha-3} = \dfrac{1}{n_\mathrm{p}} \sum \limits_{j=1}^{n_\mathrm{p}} \ln \left( \dfrac{r_j}{r_\mathrm{max}} \right) =  \dfrac{1}{n_\mathrm{p}} \sum \limits_{j=1}^{n_\mathrm{p}} \ln \left( 1 + \dfrac{r_j}{r_\mathrm{max}} -1 \right) \approx \dfrac{1}{n_\mathrm{p}} \sum \limits_{j=1}^{n_\mathrm{p}} \dfrac{r_j}{r_\mathrm{max}} - 1 \;.
\label{eq3}
\end{equation} 
The behaviour of $\rho(r)$ thus depends on the particle distribution.
The first term on the right-hand side can be interpreted as a scaled mean particle radius.
For finite $n_\mathrm{p}$ and $r_\mathrm{max}$, the upper limit is $\alpha=3$, the lower $\alpha=0$ if all $r_j \le r_\mathrm{max}$.
Now, we can explain the shape of the most common density profiles:

Let $r_\mathrm{max}=r_\mathrm{core}$ be the boundary of the core.
Debates about the slope of $\rho(r)$ in the core naturally arise because the particle number and their locations fix $\alpha$.
For a uniform particle distribution from 0 to $r_\mathrm{max}$, $\alpha =1$.
As simulations probe smaller radii, the slopes of heuristically fitted models become shallower towards $r=0$. 
This trend is explained by Equation~\ref{eq3}, when $r_\mathrm{max}$ of the probed part shrinks towards the radius of the first bin in the simulation, putting all particles at radii just below $r_\mathrm{max}$.

From now on, $r_\mathrm{max}$ is the boundary of the entire halo.
Asssuming the particle distribution becomes very dense, i.~e.~$n_\mathrm{p} \rightarrow \infty$, so that dark matter transfers \com{in}to a homogeneous fluid. 
Then, Equation~\ref{eq3} yields $\alpha=2$ for a most-likely ensemble configuration, which is interpreted as the stable, isothermal halo part every simulation and observation shows. 

The last two approximations concern the choice of $r_\mathrm{max}$, i.~e.~our definition of the extension of a halo.
Taking the limit $r_\mathrm{max} \rightarrow \infty$ and assuming that the particle farthest from the halo centre is at a much smaller, finite radius, we arrive at $\alpha=3$ belonging to a least-likely ensemble configuration. 
Depending on the choice of $r_\mathrm{max}$, shallower slopes are also possible in this approximation that omits to assign particles to the halo which are far from the halo centre but still feel its gravitational influence.

Choosing $r_\mathrm{max}$ much smaller than the extent of the particle distribution, we arrive at $\alpha=4$, assuming that, on the average, the particle radii scatter around $2\, r_\mathrm{max}$. This choice resembles models of galaxy luminosity profiles employing a half-light radius.
The respective ensemble configuration is a local log-likelihood maximum and, depending on the choice $r_\mathrm{max}$, steeper slopes are also possible.

\begin{figure*}[h!]
\begin{center}
  \includegraphics[width=0.67\textwidth]{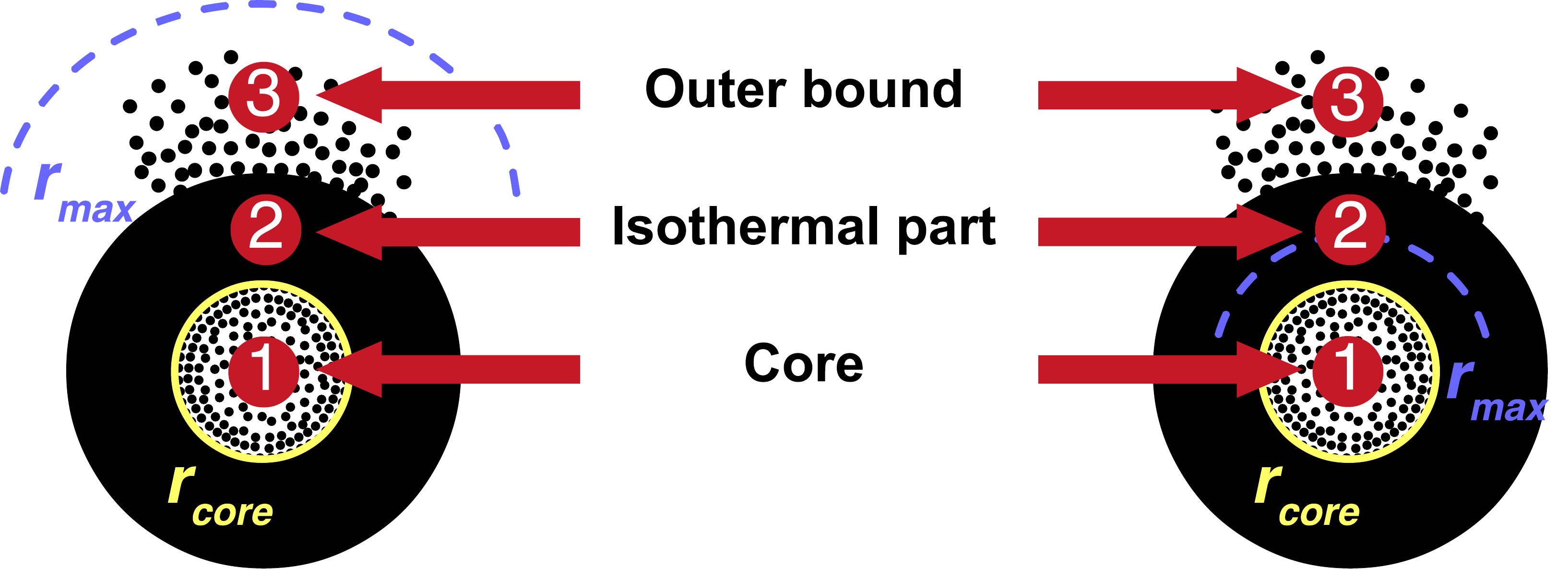}
  \caption{Monotonically  decreasing dark matter mass density profiles $\rho(r)$ for galaxy-cluster-scale halos (left) and galaxy-scale halos (right). The central region for both consists of dilute particle configurations with $\rho(r) \propto r^{-\alpha}$ and $\alpha \in \left[0, 2 \right]$ depending on the particle positions. This transfers into an isothermal part consisting of a homogeneous dark matter fluid with $\alpha=2$. At the outer bounds $\alpha \approx 3$ for galaxy-cluster-scale halos (left) and $\alpha \approx 4$ for galaxy-scale halos (right) due to the location of $r_\mathrm{max}$ relative to the particle positions.}
\label{fig:halo}
\end{center}
\end{figure*}

Considering these two boundary cases, the often found $r^{-3}$-decrease of galaxy-cluster mass densities can be explained, as well as the $r^{-4}$-decrease of galaxy mass densities. 
Due to the least- and most-likely configurations these power-law indices belong to, it is also clear that a large sample of simulated or observed galaxy clusters shows deviations from a universal behaviour with increasing resolution, while, on galaxy scale, universality may occur on average for a large amount of relaxed systems.

Summarising the results, we can decompose any dark matter halo mass density into three parts: an inner core, an isothermal region, and an outer boundary, as depicted in Figure~\ref{fig:halo}.
\\[6ex]
\noindent
\textbf{Remaining riddles}
\\[2ex]
\noindent
The approach presented here and detailed in \cite{bib:Wagner_halo} explains many dark matter halo properties found in simulations and observations in an astonishingly simple way.
It only employs a minimum amount of necessary assumptions and finally answers the question why our heuristically inferred mass density profiles are good fits to artificial and real self-gravitating dark matter structures without resorting to any cosmological model, the assembly history of the structure, or its dynamics.
Extensions to less symmetric halo shapes and the introduction of particle collisions are straightforward. 
\com{Extending the approach to alternatives of Newtonian gravity, the power-law PDF needs to be replaced to account for the characteristics of this interaction. 
We are currently investigating a derivation of $p(r)$ from more fundamental principles to be able to extend the approach to alternative theories of gravity as well.}

The approach shows that dark matter structures emerge from our halo shape definition and our findings are strongly dependent on the modelling prerequisites. 
For instance, the term ``particles" refers to the smallest constituents of the system.
In simulations, each particle is an entity of several sun masses, and, given the state-of-the-art quality of data acquisition, the same applies for observations.
It thus remains an open question how gravity and potentially other interactions shape dark matter structures beyond our current analytical, numerical, and observational limits.

A second remaining riddle is the role of $r_\sigma$. Which property of dark matter does it represent? Is it the mean free path length of actually colliding dark matter particles or \com{only} an auxiliary scaling parameter without meaning? 
Solving one mystery has entailed another. So, even in the 21st century, analysing the influence of Newton's gravity on structure morphologies remains a challenging task.
 
 \vfill

\section*{Acknowledgements}
I thank George F.~R.~Ellis and Carlo Rovelli for their inspiring works leading to this approach. In addition, I thank Xingzhong Er, Robert Grand, Jiaxin Han, Bettina Heinlein, Jens Hjorth, Sebastian Kapfer, Angela Lahee, Andrea Macciò, Christophe Pichon, Andrew Robertson, Bj\"orn Malte Sch\"afer, Johannes Schwinn, Volker Springel, R\"{u}diger Vaas, Gerd Wagner, and Liliya Williams for helpful comments, as well as the participants of the First Shanghai Assembly on Cosmology and Galaxy Formation 2019 for many helpful discussions and encouragement to further pursue this idea. I gratefully acknowledge the support by the Deutsche Forschungsgemeinschaft (DFG) WA3547/1-3.
\\[1ex]
Image credits for the ``Smiling Lens" on the cover page: NASA \& ESA

\newpage
\bibliographystyle{spmpsci}      
\bibliography{aa}   

\end{document}